\documentclass[journal]{IEEEtran}

\usepackage{algorithm}
\usepackage{algorithmic}
\usepackage{graphicx}
\usepackage[cmex10]{amsmath}
\usepackage{amssymb,amsfonts}
\usepackage[caption=false]{subfig}
\usepackage{color}
\usepackage{setspace}
\usepackage[a4paper]{geometry}
\begin{document}
\title{Topology-Age-Aware Cooperative Awareness in Vehicular Ad-Hoc Networks}

\author{\IEEEauthorblockN{Mostafa Lotfi\IEEEauthorrefmark{1} and Masoumeh Moradian\IEEEauthorrefmark{1}}

		\IEEEauthorblockA{\IEEEauthorrefmark{1}School of Computer Engineering, K. N. Toosi University of Technology, Iran}

			%	\IEEEauthorblockA{\IEEEauthorrefmark{2}Department of Electrical and Computer Engineering, Nazarbayev University, Kazakhstan}
		
		Emails:  mostafa.lotfi@email.kntu.ac.ir, mmoradian@kntu.ac.ir}

\maketitle
%\thispagestyle{empty}
%\pagestyle{empty}
%\pagenumbering{gobble}
\begin{abstract}
In vehicular ad-hoc networks (VANETs), maintaining the freshness of status information among vehicles is critical for enabling timely and reliable safety-related decisions. In this paper, we consider the problem of scheduling cooperative status broadcasting in VANETs, where vehicles are allowed to broadcast not only their own updates but also the updates of other vehicles. We propose the age-aware cooperative broadcasting (A-CB) policy, in which a road-side unit (RSU) schedules target-sender pairs based on the total achievable age-gain over the sender's network links. Moreover, through exploiting the network topology, we introduce two topology-age-aware cooperative broadcasting (TA-CB) policies; the Local TA-CB that weights links according to the number of second-hop neighbors, and the Global TA-CB that leverages the betweenness centrality of vehicles to prioritize broadcasting decisions. The proposed policies are evaluated under both random and clustered network topologies. Numerical results demonstrate that cooperative policies significantly outperform non-cooperative broadcasting. Notably, in clustered network structures, the Local and the Global TA-CB policies achieve substantial improvements over A-CB, with each outperforming the other under different network conditions.

%In this paper, we consider the problem of scheduling the cooperative broadcasting in vehicular ad-hoc networks (VANETs), where the vehicles cooperate in broadcasting the status updates of each other. We propose the age-aware cooperative broadcasting (A-CB) policy, in which the road-side-unit (RSU) schedules the vehicles for broadcasting based on the sum of age-gains achieved over their links in the network graph. We also propose local and global topology-age-aware cooperative broadcasting (TA-CB) policies that incorporate the second-hop neighbourhood structure as well as the betweenness centrality measures of the vehicles in weighing the links of each node. Through extensive numerical results, we show that the proposed cooperative policies outperform the non-cooperative one significantly. In particular, the TA-CB and GA-CB policies outperform A-CB policy in clustered network graph models.
\end{abstract}
%\vspace{-1cm}
\section{Introduction}

The freshness of information has emerged as a critical performance metric in internet-of-things (IoT)-driven technologies such as smart cities, industrial monitoring and intelligent transportation systems (ITS). Specifically, in ITS, the network structure is dynamic and the network entities, such as vehicles, should be continuously updated about the status of surrounding vehicles to make timely and appropriate safety-related decisions \cite{dl:Das}. This awareness feature is provisioned in cellular vehicle-to-everything (C-V2X) standard, where the vehicles can broadcast their status updates through cooperative aware messages (CAMs), and also issue event-triggered warnings through decentralized environmental notification messages (DENMs) \cite{dl:Chen,dl:Geeth,dl:Kueppers}. However, preserving the freshness of the broadcast information across the network remains challenging due to the dynamic environment and relies substantially on applied broadcasting policies.   

The information freshness is characterized by the age of information (AoI), which is defined as the time elapsed since the generation time of the most recent update at a monitoring node \cite{dl:Kaul}. The AoI-aware broadcasting policies have been studied in different network scenarios including centralized \cite{dl:Kadota, dl:Habob} and distributed broadcasting \cite{dl:Ni} as well as networks with fixed \cite{dl:Kadota}, and dynamic structures \cite{dl:Baiocchi,dl:Ni,dl:Jang}. 
In centralized scenarios, where a central unit is responsible for broadcasting the status updates to a set of clients, either fixed \cite{dl:Kadota}, or mobile ones such as vehicles \cite{dl:Habob}, the primary objective is to schedule the transmission and power resources in the central unit such that the expected sum AoI (ESAoI) of the clients is minimized. In the case of a one-hop downlink network with fixed clients and unreliable channels, it is shown that the greedy policy which schedules the client with the highest AoI for transmission, referred to as max-age scheduling, minimizes the ESAoI under symmetric channel conditions. On the other hand, the max-weight and Whittle's index policies outperform in nonsymmetric conditions \cite{dl:Kadota}. When the clients are moving, such as vehicles in VANETs, the age-optimal policies incorporate the vehicle positions as decision variables as well, thus, leading to more complex strategies that are not necessarily index-based, such as deep reinforcement learning (DRL)-based solutions \cite{dl:Habob}. %However, in the centralized scenario it is often assumed that the central node, which is in charge of updating the clients, has the  fresh version of the updates, 

In contrast to centralized broadcasting, distributed broadcasting involves multiple sources in the network transmitting their information to nearby entities to keep their surroundings aware of their status \cite{dl:Meng}. A typical example is cooperative awareness (CA) in modes 1 and 3 of the C-V2X standard \cite{dl:Vukadinovic,dl:Ali,dl:Xu}, where each vehicle, when scheduled by an road-side unit (RSU), broadcasts its own updates (e.g., speed and position) to surrounding vehicles. In such a case, when the RSU and all vehicles are assumed to be within the same transmission range, every entity effectively obtains the same AoI for any given vehicle \cite{dl:Elsayed}. In this setting, a max-age scheduling policy, which prioritizes the node with the highest AoI perceived at the RSU, has been proposed to minimize ESAoI \cite{dl:Jang}.
However, when vehicles are moving and not within each other's transmission range, the AoI of a specific vehicle may differ across other vehicles in the network, as it can only update a limited and varying set of neighbors whenever it is allowed to broadcast. In this case, it is proposed to prioritize those vehicles with higher ESAoI at their neighbors \cite{dl:Ni}. Nevertheless, because vehicles in a VANET are not in transmission range of each other frequently, cooperative broadcasting, where vehicles also broadcast the updates of other vehicles, can significantly improve the freshness of information across the network compared to the non-cooperative case. This cooperative approach is the main focus of this paper.

In this paper, we study centralized AoI-aware scheduling for cooperative distributed broadcasting in VANETs. We consider a scenario in which an RSU schedules vehicles for broadcasting their own information as well as the updates of other vehicles. We propose the AoI-aware cooperative broadcasting (A-CB) algorithm, in which scheduling decisions are taken based on the age-gains, i.e., the reductions in AoIs, achievable over the links of a broadcasting vehicle when it transmits the update of a specific target vehicle.
Furthermore, for the first time, we introduce topology-age-aware cooperative broadcasting (TA-CB) policies that explicitly incorporate the network graph topology into scheduling decisions. Specifically, we propose the Local TA-CB and the Global TA-CB policies. Compared to A-CB, the Local TA-CB policy takes into account the second-hop neighborhood of each broadcasting vehicle to assign weight to the vehicles. It is shown that the Local TA-CB, not only decides on instantaneous achieved age-gains but also creates additional opportunities for future AoI reduction. On the other hand, the Global TA-CB policy assigns weights based on the betweenness centrality of vehicles, a global network measure, to prioritize broadcasting decisions.
We implement our proposed policies under two network models, including random and clustered. Our results show that the cooperative policies significantly outperform the non-cooperative ones across network models. In particular, the topology-aware policies, Local TA-CB and Global TA-CB, outperform the A-CB in the clustered model, where topological diversity across different parts of the network can be exploited. Our contributions are summarized as follows:

\begin{itemize}
\item We consider a scenario in which a road-side-unit (RSU) schedules vehicles for cooperative awareness at the beginning of each frame. For such a cooperative scenario, where the vehicles cooperate in broadcasting each other's updates, we propose the A-CB algorithm that assigns indices to the pairs of broadcasting vehicles and the broadcast information based on the age-gain achieved over the outgoing links of the sender vehicle. We demonstrate that A-CB policy significantly outperforms the non-cooperative policy.

\item We propose the Local TA-CB policy, which assigns weights to broadcasting vehicles based on their potential to create opportunities for reducing the total ESAoI in upcoming frames. In particular, the number of second-hop neighbors of a broadcasting vehicle that are reachable form one of its links is exploited to weigh that link. We also propose the Global TA-CB policy which leverages the betweenness centrality measure to weigh the links, aiming to transmit updates to the distant parts of the network. 

\item Through extensive simulations, we show that all cooperative policies outperform the non-cooperative policy across various network settings. Furthermore, in clustered network structures, both the Local and the Global TA-CB achieve significant improvements over A-CB, with each policy excelling under different conditions, i.e., the Local TA-CB performing better in highly connected regimes, while the Global TA-CB excels in sparse or intermediate connectivity scenarios. These results highlight the importance of incorporating topological information into AoI-aware scheduling for VANETs. We show that, with appropriate parameter settings, the Global TA-CB can outperform the local version, albeit at the cost of higher computational complexity.
\end{itemize}

The rest of this paper is organized as follows. In Section \ref{sec:system-model}, we introduce the proposed scenario and the optimization goal. In Section \ref{sec:ACB} and \ref{sec:TA-CB}, we introduce the A-CB and the TA-CB policies, respectively. The numerical results are discussed in Section \ref{sec:numerical-results}. Finally, conclusions are drawn in Section \ref{sec:conclusion} along with potential future directions.

\subsection{Literature review}
\label{subsec:literature review}
The problem of update broadcasting in downlink one-hop networks is studied in \cite{dl:Kadota,dl:Habob}. In \cite{dl:Kadota}, the authors study a scenario in which a BTS, having the updates of a set of fixed clients at the onset of time frames, is supposed to select a subset of clients for updating. The authors show that the max-age scheduling minimizes ESAoI for symmetric clients, while the max-weight and Whittle's index policies outperform in nonsymmetric channel conditions. In \cite{dl:Habob}, the RSU is responsible for broadcasting the updates of a set of physical processes to the clients, where clients are vehicles on a road. A DRL-based algorithm is proposed to optimize the transmission mode and power of the RSU while minimizing the ESAoI of vehicles.

On the other hand, age-aware broadcasting in VANETs has also been studied in several works \cite{dl:Baiocchi,dl:Ni,dl:Jang,dl:Xu}. In \cite{dl:Baiocchi}, CA is performed in a distributed manner, where each vehicles broadcast its own updates periodically according to the IEEE802.11p standard, a random access protocol for direct short-range communications in V2X. The authors in \cite{dl:Jang} propose both centralized and decentralized AoI-based broadcast protocols for VANETs. In the centralized version, the RSU selects the vehicles with maximum perceived AoIs at RSU for broadcasting over their selected channels, assuming that all vehicles and RSU are in transmission range of each other. In \cite{dl:Ni}, the vehicles are not in the same transmission range and the RSU selects those vehicles that the weighted ESAoIs in their neighbors are larger for broadcasting, where the weights are the transmission success probabilities over links.

The randomized CA in VANETs can also be considered as a special case of gossip networks. In gossip networks, the nodes transmit their available information randomly to their neighbours with the goal of maintaining the freshest information. In \cite{dl:Kaswan1}, the authors study the version AoI in a gossip network with one information source and $n$ nodes, where the update transmission process between the source and other nodes and also among the neighbouring nodes happens according to independent Poisson processes. The information freshness in gossip networks has also been studied under various scenarios, such as the presence of unreliable sources \cite{dl:Kaswan3} and transmission mutation \cite{dl:Kaswan2}, albeit assuming age-independent spreading processes at the nodes. However, in \cite{dl:Mitra}, due to the limited updating rate in the network, each node waits for a random time proportional to its age before updating its neighbors, thus, giving  the nodes with lower AoI more opportunities to disseminate their information. In \cite{dl:Selen}, the average and variance of the AoI have been derived in gossip networks with linear, ring, and star structures. Here, the nodes transmit the information of other nodes in addition to their own information according to a time-invariant distribution. In the aforementioned works except \cite{dl:Selen}, the nodes broadcast only their own information and do not cooperate in transmitting the updates of other nodes. Furthermore, in \cite{dl:Selen}, the cooperation is done in a distributed manner, and the cooperation process is irrespective to the current AoI of the nodes. Finally, None of the above works consider centralized scheduling in gossip networks.

\section{System Model}
\label{sec:system-model}
We consider a scenario in which an RSU is responsible for scheduling the broadcast transmissions of the vehicles through its coverage area. The set of vehicles in this area is assumed to be fixed, and is indicated by $\mathcal{V}=\{1,2,\cdots,N\}$. Moreover, when allowed by RSU, each vehicle can broadcast either its own information or the information of other vehicles. The broadcast content includes vehicles' status updates such as location and speed updates as well as emergency alarms.

We assume a slotted time structure, where all vehicles are time-synchronous.  Each slot duration is equal to the time required for broadcasting one single update, which is identical for all vehicles. Moreover, every $L$ slots constitute a frame. It is assumed that the relative positions of the vehicles do not change within a frame. Furthermore, due to limited transmission power, each vehicle has a limited communication range. As a result, the vehicular network during frame $t$ is represented by the graph $G(t)=(\mathcal{V},E(t))$, where $E(t)$ indicates the edge set and $(i,j)\in E(t)$ implies that vehicles $i$ and $j$ are within each other's communication area during frame $t$ and can successfully transmit updates to each other. Moreover, $A(t)$ denotes the adjacency matrix of $G(t)$, where $A_{ij}(t)=1$ if $(i,j)\in E(t)$ and zero otherwise. 

At the beginning of each frame $t$, the RSU receives the position information of all vehicles through an uplink control channel and thus, is able to construct $G(t)$. Based on $G(t)$ and obtained AoI information of the vehicles (discussed later in this section), RSU broadcasts the vector $\mathcal{B}(t)=(\mathcal{B}_1(t),\mathcal{B}_2(t),\cdots,\mathcal{B}_L(t))$ to all vehicles, where the entry $\mathcal{B}_l(t)$ specifies the list of broadcasting vehicles in the $l$-th slot of frame $t$ as well as their broadcasting information. In particular, we refer to a broadcasting vehicle as the \emph{sender} and the vehicle whose information is being broadcast by the sender as the \emph{target}. Then, $\mathcal{B}_l(t)$ is written as  $\mathcal{B}_l(t)=\{(T_{1},S_{1}),(T_{2},S_{2}),...,(T_{m},S_{m})\}$, where the ordered pair $({T_{x}},{S_{x}}), x\in\{1,\cdots,m\}$ is an authorized \emph{target-sender} pair in slot $l$, i.e., vehicle ${S_{x}}$ broadcasts the update of vehicle ${T_{x}}$ in that slot. Whenever ${S_x}$ broadcasts the update of ${T_x}$, all neighbors of ${S_x}$, denoted by $\mathcal{N}_{S_{x}}(t)$, receive the update of vehicle ${T_{x}}$ and replace any stale information they have for vehicle $T_x$ with the newly received one. 

We also consider the following assumptions. First, we assume that any two senders scheduled to transmit in the same slot should not be neighbors or have any neighbors in common, otherwise their transmission will interfere with each other or at their common neighbors, respectively. Second, the number of broadcasting vehicles per slot is assumed to be limited, i.e., $|\mathcal{B}_l(t)| \leq M, \forall t,l$, in order to control the signaling overhead at RSU as well as the complexity of computing the target-sender pairs at the beginning of each frame. Finally, it is assumed that the update of any vehicle is broadcast only once within each frame. 
%In the scenario of this paper, we assume that the information of any vehicle in updated only once within a frame either by itself or another sender. Furthermore, the ages are updated at the end of the frames. As a result, the order of transmissions of vehicles within one single frame does not affect their ages. Hence, with a slight abuse of notation, hereafter, instead of \emph{ordered-tuple}, we refer to $B(t)$ at the \emph{set} of sender-target pairs at frame $t$. This will be discussed more in Section \ref{sec:W-ACB}.

For ease of notation, hereafter, the indices $k$, $i$, and $j$ are used for the target, sender, and receiver vehicles, respectively. The AoI of vehicle $k$ at vehicle $j$ at the beginning of frame $t$, denoted by $X_{kj}(t)$, is written as
\begin{equation}
X_{kj}(t)= t-u_{kj}(t).
\label{eq:age_def}
\end{equation}
where $t$ is the index of the current frame and $u_{kj}(t)$ is the frame index at which the most recent update of vehicle $k$ available at vehicle $j$ is generated.
 Furthermore, $X_{kj}(t)$ evolves as follows:
\begin{equation}
\begin{aligned}
X_{kj}(t+1)=\begin{cases}
X_{ki}(t)+1 ;&\exists l:  (k,i) \in  \mathcal{B}_l(t),~ j\in \mathcal{N}_i(t), \\ &X_{ki}(t)<X_{kj}(t) , \\
X_{kj}(t)+1 ;  &\text{otherwise}.
\end{cases}
\label{eq:age_evolve}
\end{aligned}
\end{equation}
%where $\mathcal{N}_j(t)$ is the neighbor set of $\nu_j$ at frame $t$, i.e., the set of vehicles in the coverage area of $\nu_j$ at frame $t$. 
The first case of \eqref{eq:age_evolve} indicates that vehicle $j$ receives an update of target vehicle $k$ in frame $t$ only if $(k,i)$ is a target-sender pair in this frame, $j$ is a neighbor of $i$, and the broadcast update by vehicle $i$ is fresher than the available one at vehicle $j$, i.e., $X_{ki}(t)< X_{kj}(t)$. Otherwise, $X_{kj}(t)$ only increases by one, as stated in the second case of \eqref{eq:age_evolve}, where one accounts for one frame duration. It is worth noting that \eqref{eq:age_evolve} is written based on the assumption that the information of any target vehicle $k$ is updated only once in a frame.

Using $X_{kj}(t)$, the ESAoI, denoted by $\bar{\Delta}$, is written as: 
\begin{equation}
\bar{\Delta}=lim_{T\rightarrow \infty }\frac{1}{T} \frac{\sum_{t=1}^{T}\sum^N_{k,j=1} \mathbb{E}[X_{kj}(t)]}{N^2}.
\label{eq:obj}
\end{equation}
In the above equation, it is assumed that the vehicles remain in the coverage area of the RSU for a sufficiently long time such that the time-average AoI can be derived regarding \eqref{eq:obj}. Moreover, the expectation is taken over the random movement of the vehicles, i.e., random $G(t)$.

Regarding that the RSU has access to $G(t)$, it is capable of tracking the changes in $X_{kj}(t), \forall k,j$, according to \eqref{eq:age_evolve}, and thus, has the updated values of AoIs at the beginning of each frame. Then, it chooses the target-sender pairs in each frame, such that the ESAoI is minimized. Consequently, the optimization problem is written as
\allowdisplaybreaks
\begin{subequations}
\label{eq:opt}
\begin{align}
&\min_{\mathcal{B}(t),\forall t} \bar{\Delta} \notag \\
\text{s.t.}\quad & k \neq k', \notag \\
&\qquad \forall t,\; \forall (k,i), (k',i') \in \cup_{l=1}^L \mathcal{B}_l(t),\; (k,i) \neq (k',i'), \label{eq:cona} \\
& \mathcal{N}_i(t) \cap \mathcal{N}_{i'}(t) = \emptyset,~~ i\notin \mathcal{N}_{i'}(t) \notag \\
&\qquad \forall l,t,\; \forall (k,i), (k',i') \in \mathcal{B}_l(t),\; i \neq i', \label{eq:conb} \\
& |\mathcal{B}_l(t)| \leq M, \quad \forall l,t, \label{eq:conc}
\end{align}
\end{subequations}
where $\mathcal{N}_i(t)$ denotes the set of neighbors of vehicle $i$ in frame $t$. Constraint \eqref{eq:cona} guarantees that each target vehicle update is broadcast at most once within a frame. Constraint \eqref{eq:conb} prohibits interfering senders, i.e., those with common neighbors, to transmit in the same slot. Finally \eqref{eq:conc} bounds the number of senders in each slot to $M$.

The above optimization problem can be formulated within an infinite-horizon Markov decision process (MDP) framework. More specifically, according to standard MDP terminology, the MDP problem is characterized by the following components:

\emph{State space}: The state is denoted by $S(t)=(G(t),\mathbf{X}(t)=[X_{ij}(t)])$. Unlike $\mathbf{X}(t)$, which has a fixed dimension of $N^2$, $G(t)$ can be represented by its edge set $E(t)$. Consequently, its dimension is of order $O(Nd)$, where $d$ is the average degree of the network. As $d$ grows to $O(N)$ in dense scenarios, the state dimension can reach $O(N^2)$, further exacerbating the curse of dimensionality.

\emph{Action space}: The action in each slot consists of selecting at most $M$ non-interfering senders as well as their corresponding targets. It is worth noting that for two vehicles to be non-interfering, their distance in graph $G(t)$ must be at least three so that they are neither adjacent nor share a common neighbor. Thus, the set of non-interfering senders in each slot corresponds to a set of vertices with pairwise distance of at least three, where such a set in graph theory is referred to as a 2-independent set of $G(t)$. Hence, the action space comprises all subsets of size at most $M$ from any 2-independent set of $G(t)$, paired with $M$ arbitrary target vehicles. However, enumerating all 2-independent sets is generally an NP-hard problem, rendering the action space prohibitively large.

\emph{Transition probabilities}: The transition probability between states is determined by the mobility pattern of the vehicles, i.e., evolution of $G(t)$ across time, as well as the changes in $\boldmath{X}(t)$, which happens according to \eqref{eq:age_evolve}.

\emph{Reward}: The reward obtained in frame $t$ is the average AoI of all vehicles, i.e., $\frac{\sum_{k,i} X_{kj}(t)}{N^2}$.

As can be seen, the MDP formulation suffers from the curse of dimensionality, arising from both the state-space size, which is of order $O(l N^3)$, and the combinatorially large action space. To address this, we propose two light-weight ESAoI-minimizing algorithms, namely, AoI-aware cooperative broadcasting (A-CB), and two Topology-AoI-aware cooperative broadcasting (TA-CB) that take advantage of age-gain minimization as well as the underlying vehicular network topology.  

\section{AoI-aware Cooperative Broadcasting (A-CB)} 
\label{sec:ACB}
In this section, we first introduce the age-gain of a target over a specific link as well as the age-gain of a target-sender pair. Then, we propose the A-CB algorithm. 

The age-gain of the target vehicle $k$ over the link $(i,j)\in E(t)$, denoted by $Q^{(k)}_{ij}(t)$, is defined as the AoI reduction of vehicle $k$ at vehicle $j$, given that vehicle $i$ broadcasts the update of vehicle $k$. Thus, $Q^{(k)}_{ij}(t)$ is written as:
\begin{equation}
Q^{(k)}_{ij}(t)=\{X_{kj}(t)-X_{ki}(t)\}^+,
\label{eq:link-gain}
\end{equation}
where $\{x\}^+=max\{0,x\}$, and is applied in \eqref{eq:link-gain} since the gain is achieved only if the update of the broadcasting vehicle $i$ is fresher than that of receiving vehicle $j$, i.e., $X_{ki}(t) \leq X_{kj}(t)$. Also, the age-gain of the target-sender pair $(k,i)$ at the beginning of frame $t$, denoted by $Q^{(k,i)}(t)$, is defined as the sum of age-gains over vehicle $i$'s links. Thus, $Q^{(k,i)}(t)$ is computed as:
\begin{equation}
Q^{(k,i)}(t)=\sum_{j\in \mathcal{N}_i(t)} Q^{(k)}_{ij}(t).
\label{eq:pair-gain}
\end{equation}

Now, we derive the sum AoI (SAoI) reduction in frame $t$, i.e.,
\begin{equation}
 \frac{1}{N^2}\sum_{ij}  X_{ij}(t)-X_{ij}(t+1),
\end{equation}
in terms of the introduced age-gains. Let us define $\mathcal{P}(t)$ as
\begin{equation}
\mathcal{P}(t)=\{(k,j)|\exists i: (k,i)\in \cup_{l=1}^{L} \mathcal{B}_l(t), i\in \mathcal{N}_j(t)\}.
\end{equation}

In fact, $\mathcal{P}(t)$ indicates all pairs of vehicles $(k,j)$, where vehicle $j$ is being updated about target vehicle $k$ during frame $t$. Then, the SAoI in frame $t$ is derived as (the fixed term $\frac{1}{N^2}$ is omitted): 

\begin{equation}
\begin{aligned}
&\sum_{k,j}  X_{kj}(t)-X_{kj}(t+1)  \overset{(a)}{=} \\
& \sum_{(k,j)\notin \mathcal{P}(t)} -1 + \sum_{(k,j)\in \mathcal{P}(t)} X_{kj}(t)-X_{kj}(t+1)\overset{(b)}{=} \\
&|\mathcal{P}(t)|-N^2+\sum_{\substack{l=1}}^L \sum_{\substack{(k,i)\\ \in  \mathcal{B}_l}} \sum_{\substack{j \\ \in  \mathcal{N}_i(t)}} \{X_{kj}(t)-X_{ki}(t)\}^+-1\overset{(c)}{=} \\
&-N^2+\sum_{l=1}^{L}\sum_{(k,i)\in \mathcal{B}_l} \sum_{j\in \mathcal{N}_i(t)} Q^{(k)}_{ij}(t)\overset{(d)}{=}\\
&-N^2+\sum_{l=1}^{L}\sum_{(k,i)\in \mathcal{B}_l}  Q^{(k,i)}(t),
\end{aligned}
\label{eq:reduction}
\end{equation}
where equality (a) is written regarding the fact that $X_{kj}(t)$ for a vehicle $j$ that is not updated about vehicle $k$ in frame $t$, i.e.,  $(k,j)\notin \mathcal{P}(t)$, increases by one, i.e., $ X_{kj}(t)-X_{kj}(t+1) =-1$.
%only the AoI of the target vehicles will change at  the neighbours of their corresponding senders at the end of frame $t$.
 Also, equality (b) is written since for $(k,i)\in \mathcal{P}(t)$, we have $X_{kj}(t+1)-X_{kj}(t)=\{X_{ki}(t)-X_{kj}(t)\}^+-1$, according to \eqref{eq:age_evolve}. Finally, equalities (c) and (d) are written using \eqref{eq:link-gain} and \eqref{eq:pair-gain}, respectively.

In the A-CB algorithm, the RSU exploits a greedy approach, i.e., based on the information about $G(t)$ and $\mathbf{X}(t)$, it selects those target-sender pairs that maximize the SAoI reduction in \eqref{eq:reduction}, or equivalently maximize $\sum_{l=1}^{L}\sum_{(k,i)\in \mathcal{B}_l}  Q^{(k,i)}(t)$. Suppose that $y^{(k,i)}({l})$ is equal to one if the target-sender pair $(k,i)$ is allowed to transmit in the $l$-th slot in frame $t$, and zero otherwise. Then, the optimization problem of A-CB in frame $t$ is written as 

\begin{subequations}
\label{eq:opt}
\begin{align}
&\max_{y^{(k,i)}(l)} \sum_{l=1}^{L} \sum_{(k,i)} y^{(k,i)}(l) \, Q^{(k,i)}(t) \notag \\
&\text{s.t.}\quad \sum_{l=1}^{L} \sum_{i=1}^{N} y^{(k,i)}(l) \leq 1, \quad \forall k \label{eq:cona} \\
&\hspace{2.4em} \sum_{(k,i)} y^{(k,i)}(l) \leq M, \quad \forall l \label{eq:conb} \\
&\hspace{2.4em} y^{(k,i)}(l) y^{(k',i')}(l) = 0, \quad  \forall l , (k,i)\neq(k',i'), d_{i,i'} \leq 2 \label{eq:conc}
\end{align}
\end{subequations}
where $d_{i,i'}$ is the distance between vehicles $i$ and $i'$ in frame $t$.
The first constraint in \eqref{eq:opt} indicates that each vehicle status is updated at most once in a frame, while the second constraint states that the maximum number of transmissions inside a slot is $M$. Also, the last constraint ensures selection of non-interfering senders inside a slot.

It is worth noting that if we set $M=1$, then \eqref{eq:conc} becomes irrelevant and the answer to \eqref{eq:opt} is to assign the $L$ sender-target pairs with maximum age-gains and distinct targets to $L$ slots in each frame. However, if $M>1$, then \eqref{eq:conc} comes to the picture and the optimization \eqref{eq:opt} becomes a quadratically constrained integer program, which requires the information of the graph structure $G(t)$ to establish the quadratic constraints in \eqref{eq:conc}. These constraints increase the complexity of solving the problem, especially in the case of connected and dense graphs. In this regard, we propose A-CB policy in Algorithm \ref{alg:ACB} to solve \eqref{eq:opt}. 

%In A-CB policy, first, the pool of eligible target-sender pairs for each slot $l$, referred to as $Pool_l, \forall l$, is initialized to include all possible sender-target pairs (line 1). Also, a list $\mathcal{S}$ is constructed which sorts all possible sender-target pairs in a descending order according to their $Q^{(i,k)}$ values (line 2). Then in each iteration, the first element of the list $\mathcal{S}$, say for example $(i,k)$, is selected (line 5), and is assigned to the first slot $l$ that satisfies two conditions; First $(i,k)$ is eligible in slot $l$, i.e., $(i,k) \in Pool_l$, and second, the capacity $M$ is not reached in that slot (lines 8,9). If such a slot $l$ exists, all sender-target pairs with target $k$ are removed from all pools (line 10) since any target is allowed to be updated only once in each frame. Moreover, the pairs with interfering senders $j \in \mathcal{N}_i(t)$ are removed from $Pool_l$ (line 11). If the pair $(i,k)$ cannot be assigned to any slot, it is removed from the head of $\mathcal{S}$ (lines 16-18). This process continues all slots are either reached to their capacity or their pool are empty, or list $\mathcal{S}$ becomes empty (i.e., conditions in line 4).  

In the A-CB policy, a pool of eligible target-sender pairs is first initialized for each slot $l$, denoted as $Pool_l, \forall l$, containing all possible sender-target pairs (line 1). Additionally, a list $\mathcal{S}$ is constructed, which sorts all possible target-sender pairs in descending order of their $Q^{(k,i)}$ values (line 2). Then, in each iteration, the first element of $\mathcal{S}$, say $(k,i)$, is selected (line 5) and assigned to the first slot $l$ that satisfies two conditions. First, $(k,i)$ is eligible in slot $l$, i.e., $(k,i) \in Pool_l$; and second, the capacity $M$ of slot $l$ has not yet been reached (lines 8,9). If such a slot exists, all target-sender pairs involving target $k$ are removed from all pools, because each target is allowed to be updated only once per frame (line 10). Moreover, pairs with interfering senders, i.e., $\forall i': d_{i,i'}\leq 2$ are removed from $Pool_l$ (line 11). If the pair $(k,i)$ cannot be assigned to any slot, it is removed from the head of $\mathcal{S}$  (lines 16-18). This process continues until every slot either has reached its capacity or has an empty pool, or until the list $\mathcal{S}$ becomes empty (i.e., the conditions in line 4 are met).

\begin{algorithm}[!t]
\caption{A-CB Algorithm}
\label{alg:ACB}
\begin{algorithmic}[1]
\REQUIRE $\forall (k,i):Q^{(k,i)}(t)$, $G(t)$, $M$, $L$
\ENSURE Set of target-sender pairs assigned to each slot
\STATE Initialize $Pool_l = \mathcal{V} \times \mathcal{V}$ for $1 \le l \le L$ \COMMENT{eligible pairs per slot}
\STATE $\mathcal{S} \gets$ sorted list of $\mathcal{V} \times \mathcal{V}$ in descending order of $Q^{(i,k)}$
\STATE $n(l) \gets 0$ for $1 \le l \le L$ \COMMENT{number of assigned pairs in slot $l$}
\WHILE{$\bigl(\exists l : Pool_l \neq \emptyset \;\land\; n(l) < M\bigr) \;\land\; (\mathcal{S} \neq \emptyset$)}
    \STATE $(k,i) \gets \mathcal{S}[1]$ \COMMENT{head of sorted list}
    \STATE $flag \gets 0$
    \FOR{$l = 1$ to $L$}
        \IF{$(k,i) \in Pool_l$}
            \STATE Assign $(k,i)$ to slot $l$
            \STATE $n(l) \gets n(l) + 1$
            \STATE Remove all $(k,i')$ from $Pool_{l'}$ for every $l'$ \COMMENT{no other sender can send update of vehicle $k$}
            \STATE Remove all $(k',i')$ with $ d_{i,i'} \leq 2$ from $Pool_l$ \COMMENT{interference constraint}
            \STATE $flag \gets 1$
            \STATE \textbf{break}
        \ENDIF
    \ENDFOR
    \IF{$flag = 0$}
        \STATE Remove $(k,i)$ from the head of $\mathcal{S}$
    \ENDIF
\ENDWHILE
\end{algorithmic}
\end{algorithm}

%\begin{figure}[!t]  % placement: here, top, bottom, page
%\caption{Example: A-CB Algorithm}
%\begin{algorithmic}[1]
%\label{alg:ACB}
%\textbf{Input}: $[Q_{ij}(t)]$, $G(t)$, $M$, $L$
%\textbf{Output}: sender-target pairs
%\State Set $\mathcal{S}(i) = \mathcal{V}, ~~\forall i\in \{1,2,\cdots,L\}$
%\end{algorithmic}
%\end{figure}

%Please note that when $\nu_i$ broadcasts the update of $\nu_k$ in frame $t$, the ages $X_{kj}(t),~ j\in \mathcal{N}_i(t)$, and consequently, the age-gains $Q^{(k)}_j, j\in {\mathcal{N}_i}$ will change, However, the age-gain of other sender-target pairs $(i,k')$ with $k'\neq k$ remain the same. Thus, setting $y_{(i,k)}=1$ does not change the values of $Q^{(k')}_{j}, ~ k'\neq k, \forall j$. In this regard, the optimal solution of \eqref{eq:opt} is to choose the $L$ largest members of the following set:
%\begin{equation}
%\{(s_k , k)| s_k = \text{argmax}_{i} ~Q^{(k)}_i\},
%\end{equation}
  
\section{Topology-AoI-Aware Cooperative Broadcasting (TA-CB)}
\label{sec:TA-CB}
In calculating the age-gain of a target-sender pair, the A-CB algorithm assigns equal weights to the age-gains associated with a sender's link, as in \eqref{eq:pair-gain}. However, the neighbors of a sender have different structural positions in the network and thus, can affect the dissemination of the information differently in consequent frames.
In this regard, the main idea in the TA-CB algorithm is to take advantage of the topological position of the vehicles in the network to create more opportunities for fresh information dissemination in the upcoming frames. We propose two versions of TA-CB, namely the Local TA-CB and the Global TA-CB, where the former relies on local topological information of each vehicle, while the latter exploits the global network properties.

\subsection{Local TA-CB}
\label{subsec:local_TACB}

\begin{figure}[t!]
\centering 
\includegraphics[width=\columnwidth]{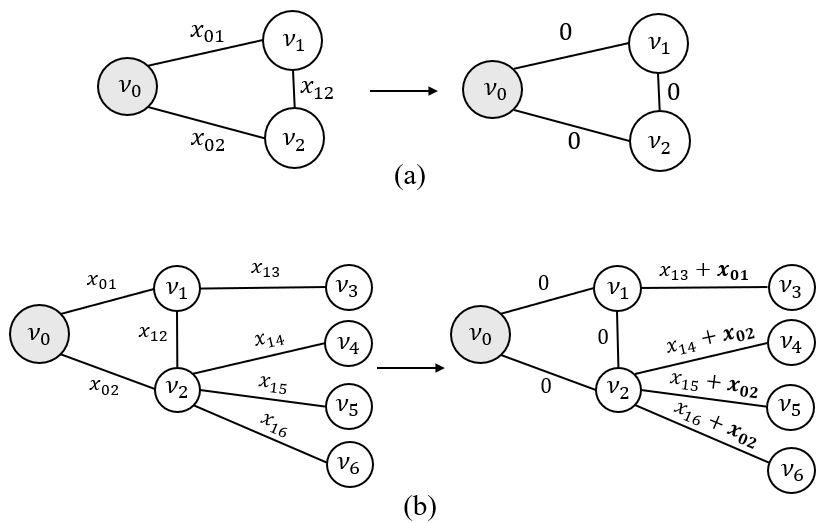}
\caption{In graphs of (a) and (b), $\nu_0$ broadcasts the update of target vehicle $k$ to its neighbors, where $x_{ij}=Q^{(k)}_{ij}$ is the age-gain of link $(i,j)$ with respect to target $k$. In (a), vehicle $\nu_0$ has no second-hop neighbors. In (b), the age-gains are added to the gains of the second-hop links. }
\label{fig:TACB}
\end{figure}

 In order to clarify the idea of the Local TA-CB, we bring an example. Consider Fig.s \ref{fig:TACB}(a) and (b) that illustrate two different neighborhoods of sender vehicle $0$, respectively. In this figure, the link age-gains are shown for a specific target vehicle $k$. Consequently, $Q^{(k)}_{ij}$ is denoted by $x_{ij}$ and is shown over the link $(i,j)$. It is also assumed that vehicle $0$ has fresher update than its neighbour vehicles $1$ and $2$, i.e., $x_{01},x_{02}>0$. Moreover, $G(t)$ is assumed to remian unchanged in two consecutive frames. 

In both scenarios in Fig.s \ref{fig:TACB}(a) and (b), when vehicle $\nu_0$ broadcasts the update of vehicle $k$, the associated age-gains over its own links as well as the links between its neighbors are reset to zero, as all neighbors obtain the same update of vehicle $k$. However, in Figure \ref{fig:TACB}(b), where vehicle $0$ has second-hop neighbors, i.e., vehicles $\{3,4,5,6\}$, the age-gains of the second-hop links also change. As such, the age-gain of the link $(1,3)$ is incremented by $x_{01}$, while the age-gains of links $(2,4)$, $(2,5)$, and $(2,6)$ are all incremented by $x_{02}$. Thus, when a sender broadcasts an update, the age-gain of each of its links are added to the corresponding second-hop links, thus, incrementing the age-gains over these links. This phenomenon of age-gain spreading provides the opportunity of having higher target-sender age-gains in frequent frames, and thus, increasing the chance of obtaining larger AoI reductions, given that the network does not change significantly in near future. Note that, the A-CB algorithm does not differentiate between the structures in Fig.s \ref{fig:TACB} (a) and (b) when computing the age-gain of the target-sender pair $(k,0)$ since in both cases we have $Q^{(k,0)}=x_{01}+x_{02}$. 

In this regard, we propose the Local TA-CB algorithm which takes into account the second hop links of each sender to determine the broadcasting target-sender pairs in each frame. 
More precisely, according to the Local TA-CB algorithm, the age-gain of target-sender pair $(k,i)$ is re-defined as:
\begin{equation}
Q^{(k,i)} = \sum_{j\in \mathcal{N}_i} w_{ij}(t) Q^{(k)}_{ij}(t),
\label{eq:sender-target-gain-w}
\end{equation}
where $Q^{(k)}_{ij}(t)$ is defined as in \eqref{eq:link-gain}. Also, the weight $w_{ij}(t)$ associated to link $(i,j)\in E(t)$ is defined as 
\begin{equation}
w_{ij}(t) = 1+N^{(2)}_{i\rightarrow j}(t)
\label{eq:weight}
\end{equation}
where $N^{(2)}_{i\rightarrow j}(t)$ is the number of second-hop neighbors of vehicle $i$ that are neighbors of vehicle $j$ as well. In other words, $N^{(2)}_{i\rightarrow j}(t)=|\mathcal{N}^{(2)}_{i\rightarrow j}(t)|$, where $\mathcal{N}^{(2)}_{i\rightarrow j}(t)$ is the set of second-hop neighbors of vehicle $i$ via vehicle $j$:
\begin{equation}
\mathcal{N}^{(2)}_{i\rightarrow j}(t)=\{m\in \mathcal{V}|m \in \mathcal{N}_j(t), m \notin \mathcal{N}_i(t)\}.
\end{equation}

It is worth noting that regarding \eqref{eq:sender-target-gain-w} and \eqref{eq:weight}, we have 
\begin{equation}
Q^{(k,i)}=\sum_{j\in \mathcal{N}_i} Q^{(k)}_{ij}(t)+\sum_{j\in \mathcal{N}_i} N^{(2)}_{i\rightarrow j}(t) Q^{(k)}_{ij}(t)
\end{equation}
where the first term is the age-gain achieved for target vehicle $k$ in the current frame as a result of having vehicle $i$ as the sender. Furthermore,  $N^{(2)}_{i\rightarrow j}(t) Q^{(k)}_{ij}(t)$ in the second term is the improvement in $Q^{(k,j)}$, since as illustrated in Fig. \ref{fig:TACB}, the age-gain $Q^{(k)}_{ij}$ is added to the age-gain of all links between $j$ and the vehicles in $ \mathcal{N}^{(2)}_{i\rightarrow j}(t)$.
Thus, the Local TA-CB prioritizes vehicles that not only achieve higher age-gains in current frame but also lead to higher age-gain improvements in subsequent frames. Albeit, this assumption is justified by the fact that the graph $G(t)$ remains unchanged over multiple frames, which is realistic in practical vehicular scenarios.

 The optimization of the Local TA-CB is written similarly to \eqref{eq:opt} except that $Q^{(i,k)}(t)$ is defined according to \eqref{eq:sender-target-gain-w}. Moreover, the algorithm of the Local TA-CB is exactly as Algorithm \ref{alg:ACB}, except that
in line 2, the list $\mathcal{S}$ is sorted according to $Q^{(i,k)}(t)$ in \eqref{eq:sender-target-gain-w}.  

\subsection{Global TA-CB}
\label{subsec:global-TACB}
In the Global TA-CB algorithm, rather than weighting a sender's links based on their second-hop neighbors as in \eqref{eq:weight}, we consider the capability of vehicles to effectively transfer the information to more distant parts of the network. In network analysis terminology, this property is captured by the betweenness centrality measure. Accordingly, the centrality of vehicle $j$ in frame $t$, denoted by $B_j(t)$, is defined as \cite{dl:newman}:
\begin{equation}
B_j(t) = \sum_{(x,y)\in \mathcal{V}^2}  \frac{n_{sp}(j,(x,y))}{n_{sp}(x,y)},
\end{equation}  
where $n_{sp}(x,y)$ is the total number of shortest paths between vehicles $x$ and $y$ and $n_{sp}(j,(x,y))$ is the number of those paths that pass through node $j$. In fact, the more shortest paths that traverse node $j$, the more critical its role in connecting different parts of the network. Consequently, we define $w_{ij}(t)$, as follows:
\begin{equation}
w_{ij}(t)=1+\alpha B_{j}(t).
\label{eq:bet}
\end{equation}

In the above equation, each link weight is inflated by $\alpha B_{j}(t)$ compared to A-CB, thereby giving more weight to the links that terminate at centrally located vehicles. Also, as $\alpha$ increases the importance of transferring information to other parts of the network becomes more than reducing  the age over local link $(i,j)$. Unlike the Local TA-CB method, the Global TA-CB method requires the entire graph structure to compute $B_{j}(t)$ and thus, is computationally more expensive. In Section \ref{sec:numerical-results}, we show that each of these two policies outperforms the other under certain circumstances. 

\section{Numerical Results}
\label{sec:numerical-results}
\subsection{Simulation setup}
\label{subsec:sim_setup}
In this section, we evaluate our proposed algorithms A-CB, Local TA-CB, and Global TA-CB against each other and an age-aware non-cooperative broadcasting policy, referred to as N-CB \cite{dl:Ni}, under different network settings. N-CB behaves similarly to A-CB, except that the vehicles only broadcast their own information. Moreover, the graph $G(t)$ is generated according to two models; First, the standard Erdos-Renyi (ER) model, in which each link is independently created according to a fixed probability $p_l$. Second, a clustered model in which $N$ vehicles are partitioned into $N_c$ equal-sized non-overlapping consecutive clusters along the road, with intra-cluster connections constructed according to the ER model. Additionally, each cluster is connected to the subsequent cluster via a single inter-cluster link, simulating the platoon-like structures. Also, in order to model realistic mobility patterns, the graph topology $G(t)$ is updated every $N_g$ frames. The default values of parameters are considered as in Table \ref{tab:default_params}, unless otherwise stated.

\begin{table}[t]
\centering
\caption{Default simulation parameters.}
\label{tab:default_params}
\footnotesize
\begin{tabular}{|c|c|}
\hline
\textbf{Parameter} & \textbf{Default Value} \\ \hline
Number of vehicles, \(N\) & 20 \\ \hline
Number of slots per frame, \(L\) & 4 \\ \hline
Maximum senders per slot, \(M\) & 2 \\ \hline
Number of clusters, \(N_c\) & 3 \\ \hline
Graph update interval, \(N_g\) & 10 \\ \hline
$\alpha$ in \eqref{eq:bet} & 1 \\
\hline
\end{tabular}
\end{table}

\begin{figure}[t]
\centering
\includegraphics[width=\columnwidth]{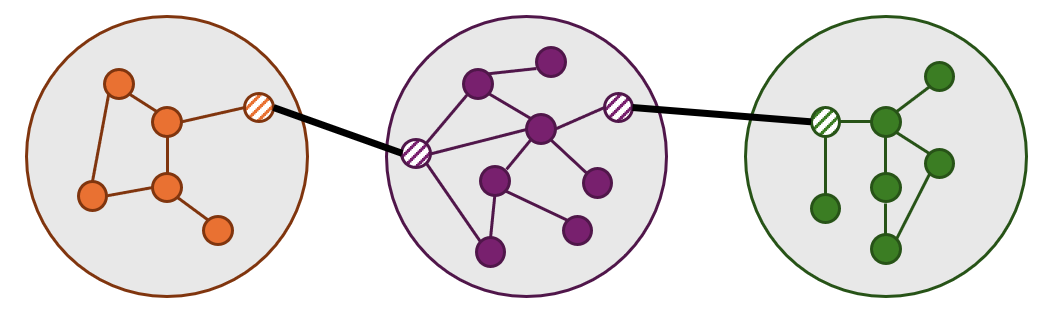}
\caption{An example of clustered graph with three clusters. The patterned circles show the connecting vehicles.}
\label{fig:cluster-graph}
\end{figure}

In our simulations, we mainly focus on the clustered structure since, as will be seen later, the TA-CB policies outperform A-CB under such topologies. Fig. \ref{fig:cluster-graph} illustrates an example of a clustered graph with three clusters, where the vehicles that are serving as the endpoints of the inter-cluster links are referred to as connecting vehicles. In the following, we analyze the effect of $p_l$, i.e., the probability of intra-cluster link construction, on the weights $w_{ij}(t)$ in the Local and the Global TA-CB policies, respectively, as this analysis is useful for interpreting the simulation results presented later. We denote the size of $c$-th cluster by $n_c$.

Consider a link $(i,j)\in E(t)$, where vehicle $i$ is the sender. Also, suppose that neither $i$ nor its neighbor $j$ is a connecting vehicle. Then, the number of second-hop neighbors of $i$ via $j$, i.e., $N^{(2)}_{i\rightarrow j}(t)$ in \eqref{eq:weight}, follows a binomial distribution $\text{Bin}(n_c-2,p_l(p_l-1))$. This is due to the fact that any other vehicle in the same cluster is a neighbor of $j$, or equivalently a second neighbor of $i$, with probability $p_l$, and is not a neighbor of $i$, with probability $1-p_l$. Thus, the expected value of $w_{ij}(t)$, defined in \eqref{eq:weight}, is 
\begin{equation}
\mathbb{E}[w_{ij}(t)]= 1+(n_c-2) p_l (1-p_l).
\label{eq:Ew}
\end{equation}

It is concluded from \eqref{eq:Ew} that  $\mathbb{E}[w_{ij}(t)]$ increases with $p_l$ initially but eventually decreases since the cluster becomes densely connected, approaching to a complete graph, as $p_l$ increases. Now suppose that link $(i,j)$ is an inter-cluster link, i.e., vehicle $i$ belongs to cluster $c$ and vehicle $j$ belongs to cluster $c+1$. In this case, $N^{(2)}_{i\rightarrow j}(t)$ follows binomial distribution $\text{Bin}(n_{c+1}-1,p_l)$ since any vehicle in the next cluster can be a second-hop neighbor of vehicle $i$ if it is a neighbor of $j$ (with probability $p_l$). Thus, the expected weight becomes
\begin{equation}
\mathbb{E}[w_{ij}(t)]= 1+(n_c-1) p_l,
\end{equation}
which increase monotonically with $p_l$. Finally, if $i$ is not a connecting vehicle but $j$ is, we have $N^{(2)}_{i\rightarrow j}(t) =1 $ and thus:
\begin{equation}
\mathbb{E}[w_{ij}(t)]= w_{ij}(t)=2.
\end{equation}

Hence, it can be concluded that in the Local TA-CB policy, for equal age-gains, the connecting vehicles are assigned higher weights due to their inter-cluster links. In particular, when $p_l$ is larger, this weight disparity becomes more noticeable.

On the other hand, in the Global TA-CB policy, the connecting vehicles possess the largest betweenness centrality values since any path from their cluster to the subsequent clusters must pass through them. Moreover, when $p_l$ is small, the clusters are sparse, and the non-connecting vehicles inside intermediate clusters exhibit higher betweenness centrality compared to those in peripheral clusters. This is because they lie on paths between the two connecting vehicles of their cluster. However, as $p_l$ increases, the clusters become more densely connected, and thus, the betweenness centralities of vehicles inside the intermediate clusters decrease as the direct link between two connecting vehicles of their clusters is more likely available. In particular, in the extreme case of $p_l=1$, all vehicles except the connecting ones have zero betweenness centrality, since the graph inside each cluster is complete and the two connecting vehicles are directly connected.% The following figure focus on the improvement on local and global TA-CB compared to A-CB under different network condition.

\subsection{Simulation results}

\begin{figure}[t]
\centering
\includegraphics[width=\columnwidth]{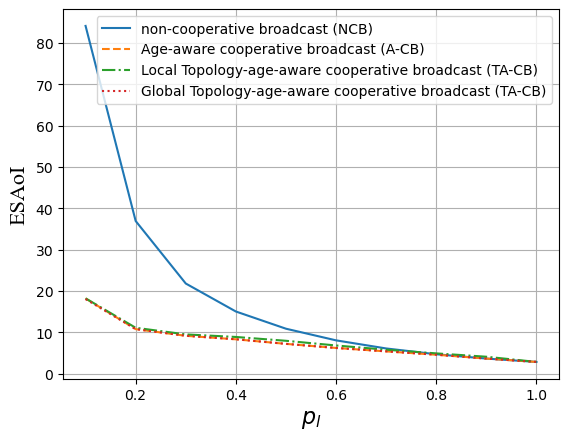}
\caption{The ESAoI versus the link construction probability, $p_l$, in random model ($L=4, M=2, N_g = 10$).}
\label{fig:random_plink}
\end{figure}

Fig. \ref{fig:random_plink} depicts the ESAoI versus the link probability $p_l$ for the random topology, under the N-CB, A-CB, Local TA-CB, and Global TA-CB policies. As observed, in cooperative policies A-CB and TA-CB, the ESAoI is significantly lower than non-cooperative policy N-CB since cooperation enables vehicles to transmit each other's updates, thereby disseminating status information more broadly across the network and reducing the ESAoI. Moreover, under all policies, ESAoI decreases with $p_l$ since higher connectivity increases the likelihood that vehicles lie within each other's communication range, and thus, receive fresher updates of each other. Furthermore, the Local and Global TA-CB do not perform better than the A-CB in the random graph since the TA-CB policies rely on the topological differences of vehicles to act, while in the random graph all vehicles have the same topological position.

\begin{figure}[t]
\centering
\includegraphics[width=\columnwidth]{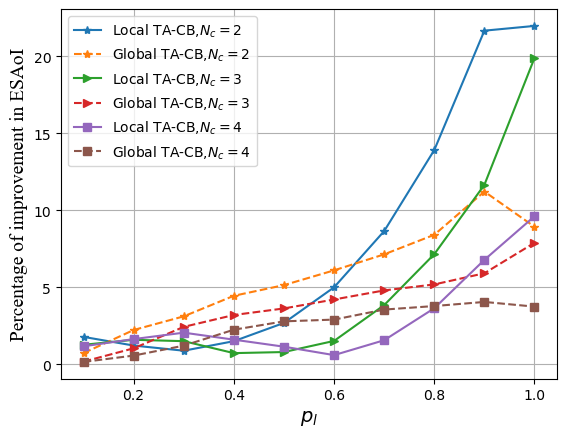}
\caption{The improvement in ESAoI versus the intra-cluster link construction probability, $p_l$, at different values of $N_c$ ($L=4, M=2, N_g=10$).}
\label{fig:clustered_plink_Nc}
\end{figure}

In Fig. \ref{fig:clustered_plink_Nc}, the percentage of ESAoI improvement in the Local and Global TA-CB over A-CB policy are shown versus the intra-cluster link probability $p_l$, for different numbers of clusters $N_c$. It is observed in Fig. \ref{fig:clustered_plink_Nc} that in general, the Local and Global TA-CB policies outperform A-CB at all values of $p_l$. However, the improvement achieved by the Local TA-CB remains almost low, i.e., less than $3$ percent, when $p_l$ is less than a certain threshold and increases with $p_l$ beyond that point. In particular, this threshold is approximately 0.3, 0.5, and 0.6, for $N_c=2,3$, and $4$, respectively. 
In fact, when $p_l$ is small, as concluded in Section \ref{subsec:sim_setup}, the inter-cluster links are not significantly prioritized over other links. Besides, the network remains sparse within clusters. 
Thus, even if an update is forwarded to other clusters via inter-cluster links, those clusters cannot effectively propagate information further due to their sparse internal connectivity. However, when $p_l$ increases beyond the threshold, the inter-cluster links receive higher weights and the clusters become sufficiently connected to spread the information effectively. Consequently, the improvement increases with $p_l$.
 On the other hand, the Global TA-CB policy exhibits a steady increase in improvement with $p_l$, even at low values. This is because in the Global TA-CB, when $p_l$ is small, the vehicles in intermediate clusters have higher betweenness centrality than those in peripheral clusters, as noted in Section \ref{subsec:sim_setup}. Thus, they are prioritized for update broadcasting, leading to a steady increase in improvement with $p_l$. Also, as $p_l$ becomes larger, though the priority of intermediate vehicles diminishes, the increased connectivity of the network compensates for this effect and the outperformance continues to grow.  

It is also observed in Fig. \ref{fig:clustered_plink_Nc} that for an intermediate range of link probabilities $p_l$, the Global TA-CB outperforms the Local TA-CB. This is due to the reason that as explained earlier, the Global TA-CB policy assigns higher betweenness centrality to intermediate vehicles at low $p_l$, thereby prioritizing them for status updates. At higher values of $p_l$, however, the Local TA-CB significantly outperforms the global one. In fact, at these values, both policies essentially treat intra-cluster links across different clusters similarly, i.e., give lower weight to them compared to inter-cluster links, which is due to the similar topological structure of the vehicles within clusters in the Local TA-CB, and low betweenness centrality in the Global TA-CB. Nevertheless, the inter-cluster links in the Local TA-CB policy are weighted significantly higher than in the Global TA-CB since the weight include the number of second-hop neighbors which is large for inter-cluster links in connected regime of the network. %As such, the difference between link weights in local TA-CB is $p_l(n_c-2)$ on average, while in the global TA-CB is the betweenness measure of the connecting vehicles. 
Consequently, the connecting links are selected more frequently for broadcasting, leading to greater improvement in the Local TA-CB. As an example the improvement is $9\%$ and $18\%$ higher compared to the Global TA-CB at $p_l = 0.9$ and $p_l=1$, respectively.

Another key observation in Fig. \ref{fig:clustered_plink_Nc} is that as the number of clusters $N_c$ increases, the improvement of both the Local and Global TA-CB over A-CB degrades. This is due to the fact that, given a fixed $N$, the number of vehicles in each cluster decreases, thus, the topological distinctiveness between vehicles within clusters and the connecting ones diminishes, making the weights $w_{ij}(t)$ less useful in the Local TA-CB. Also, in the Global TA-CB, the betweenness centrality of the connecting vehicles approaches that of the internal cluster vehicles, leading to degradation of the performance.  

\begin{figure}[t]
\centering
\includegraphics[width=\columnwidth]{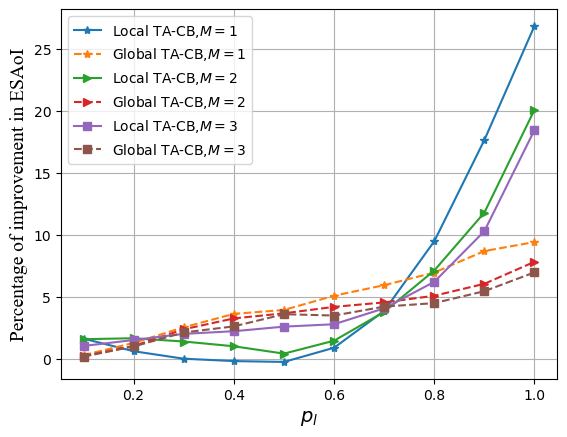}
\caption{The improvement in ESAoI versus the intra-cluster link construction probability, $p_l$, at different values of $M$ ($L=4, N_c=3, N_g=10$).}
\label{fig:clustered_plink_ntarget}
\end{figure}

Fig. \ref{fig:clustered_plink_ntarget} depicts the improvement in the Local and Global TA-CB over A-CB in the clustered model, for different numbers of transmissions per slot, $M$. 
In general, the ESAoI decreases with $M$ under all policies and across all values of $p_l$ since more broadcasting opportunities are provided in each slot, though marginal improvement diminishes as $M$ increases due to the interfering senders constraint in \eqref{eq:conc}. 
However, the amount of improvement of the Global TA-CB over A-CB, as well as that of the Local TA-CB for $p_l \geq 0.7$, decreases with $M$. This is because as $M$ increases, the base-line policy A-CB also benefits from the additional broadcasting opportunities, thus narrowing the performance gap that TA-CB policies can achieve. The opposite behaviour is observed for the Local TA-CB policy when $p_l \leq 0.7$, where the improvement grows with $M$, which is justified as follows. At low $p_l$, the network is sparse and the number of second-hop neighbors is very limited. As a result, any broadcasting primarily benefits its first-hop neighbors, without increasing the age-gains of other links (please see Fig. \ref{fig:TACB}) that could otherwise be exploited later to further reduce ESAoI as $M$ increases. In this regime, the Local TA-CB can achieve greater improvement over A-CB as $M$ increases since it prioritizes vehicles with more second-hop neighbors.

\begin{figure}[t]
\centering
\includegraphics[width=\columnwidth]{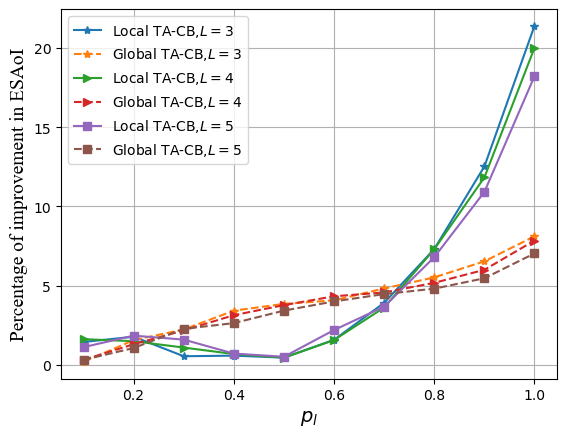}
\caption{The improvement in ESAoI versus the intra-cluster link construction probability, $p_l$, at different values of $L$ ($N_c = 3, M=2, N_g=10$).}
\label{fig:clustered_plink_Nslot}
\end{figure}

\begin{figure}[t]
\centering
\includegraphics[width=\columnwidth]{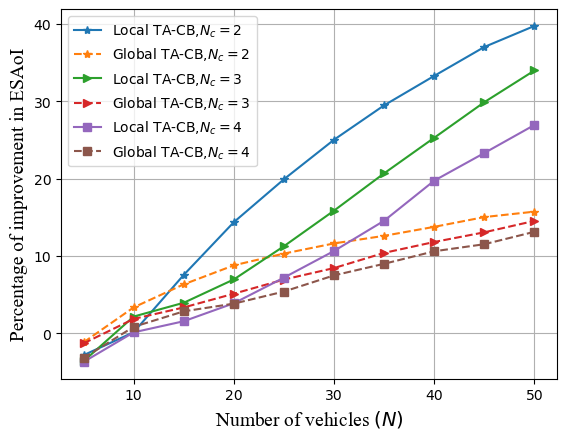}
\caption{The improvement in ESAoI versus the number of vehicles, $N$, in the clustered model, at different values of $M$ ($L=4, N_c=3, N_g=10, p_l=0.8$).}
\label{fig:clustered_Vehicle_Nc}
\end{figure}

In Fig. \ref{fig:clustered_plink_Nslot} the improvement in the Local and Global TA-CB over A-CB is shown versus $p_l$ for different number of frame lengths, $L$. The  behaviour of curves is similar to those observed in Fig. \ref{fig:clustered_plink_ntarget}. Here, increasing $L$, similar to increasing $M$, provides more broadcasting opportunities. Thus, all behaviours can be justified similar to those in Fig. \ref{fig:clustered_plink_ntarget}. The only difference is that here, the improvements at different values of $L$ are closer to each other, e.g., at $p_l=0.9$, all percentages lie within the range of $11\%$ and $12.5\%$. The reason is that, while increasing $M$ introduces additional senders within the same slot, which must satisfy the interfering-sender constraint, increasing $L$
adds new slots that are not subject to this constraint. As a result, the additional opportunities provided by larger $L$ can be more effectively exploited, leading to sustained improvement and closer performance across different frame lengths.

%both proposed algorithms, the Local and Global T-ACB, consistently outperform the baseline ACB policy across all evaluated scenarios. The figure shows that the improvement percentage in ESAoI initially shows an increasing trend, achieving its highest performance at a 25 number of vehicles as network density, after which it begins to decline. To explain this phenomenon, it is important to note that during interval of lower number of vehicles (as seen in the initial phase of the graph), the number of nodes within each cluster is at its lowest, leading to a highly sparse network. In such scenarios, the physical topology does not align with the formal definition of a clustered network, making clustering concepts ineffective. Therefore, while the proposed T-ACB algorithms still continue to outperform ACB algorithm, they are unable to entirely leverage their optimization capabilities to reach higher improvements. On the other hand, as the number of vehicles goes beyond the optimal threshold (Here equals to 25), the improvement percentage in ESAoI for the Local and Global T-ACB decreases. This decline is due to the fact that while the frame length L remains constant throughout the experiment, the number of vehicles continuously increases, leading to a highly dense network relative to L. Therefore, vehicles suffer from reduced channel access time and have fewer transmission opportunities in each frame, resulting in a lower percentage of improvement in ESAoI for the proposed algorithms.

Fig. \ref{fig:clustered_Vehicle_Nc} illustrates the percentage of improvement of the Local and Global TA-CB policies over A-CB versus the number of vehicles, $N$, for different number of clusters, $N_c$, at $p_l=0.8$. When $N$ is small (less than 10), the size of clusters is too small to exhibit a meaningful clustered structure, and thus, the TA-CB policies make little or no improvement over A-CB. However, as $N$ increases, the clusters become more crowded, enhancing the importance of inter-cluster links in disseminating information. These connecting links gain more second-hop neighbors in local TA-CB and larger betweenness centrality measures in the Global TA-CB, both of which play a decisive role in their respective policies. Also, since $p_l=0.8$ corresponds to a densely connected regime, the local TA-CB policy transfers updates more effectively across clusters, consistent with the observations made earlier.

\begin{figure}[t]
\centering
\includegraphics[width=\columnwidth]{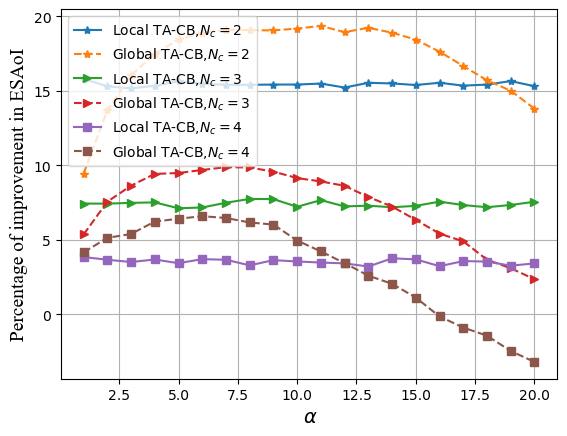}
\caption{The improvement in ESAoI versus coefficient $\alpha$ in the clustered model, at different values of $M$ ($L=3, N_g=10$, $p_l=0.8$).}
\label{fig:clustered_alpha_Nc}
\end{figure}

In Fig. \ref{fig:clustered_alpha_Nc}, the effect of $\alpha$ in \eqref{eq:bet} is examined on the improvement of the Global TA-CB policy over A-CB, at $p_l=0.8$. The improvement of local TA-CB, which is independent of $\alpha$ is also shown for the sake of comparison. It is observed in this figure that as $\alpha$ increases from one, the Global TA-CB improvement surpasses that of local TA-CB. This is due to the fact that the links ending at high-centrality vehicles are weighted more, resulting in faster information dissemination across clusters. However, as $\alpha$ exceeds a threshold, the improvement degrades. This is because transferring information to distant parts of the network are so highlighted that the current age-gain reduction over first-hop links in the current broadcast, i.e., equivalent to the first term in \eqref{eq:bet}, are overlooked. This leads to not effectively transmitting the fresher information. Also, it is worth noting that at the optimal value $\alpha=10$ at $N_c=2$, the Global TA-CB outperforms the Local TA-CB by approximately $4\%$, albeit at the cost of higher computational complexity for centrality computation.
 
\section{Conclusion}
\label{sec:conclusion}
In this paper, we studied the problem of AoI-aware scheduling for cooperative distributed broadcasting in VANETs, where vehicles cooperate by broadcasting not only their own status updates but also those of other vehicles. We considered a centralized framework in which an RSU schedules target-sender pairs at the beginning of each frame based on the network topology and the current AoI information of the vehicles. We proposed three scheduling policies. First, the A-CB algorithm, which prioritizes target-sender pairs based on the total age-gain achieved over the links of each sender. Second, the local TA-CB policy, which additionally accounts for the second-hop neighborhood of each sender to weigh its links, thereby creating more opportunities for future AoI reduction. Third, the Global TA-CB policy, which leverages the betweenness centrality of vehicles to prioritize links that terminate at structurally important nodes, facilitating faster information dissemination across the network. Through extensive simulations under random and clustered topologies, we demonstrated that cooperative policies significantly outperform non-cooperative N-CB, and that TA-CB policies achieve substantial gains over A-CB in clustered networks, with local TA-CB excelling in densely connected regimes and the Global TA-CB in sparse or intermediate connectivity scenarios, though at the cost of higher computational complexity for centrality computation. This highlights the importance of topology-aware scheduling in VANETs and the trade-off between performance and complexity. Future research directions include distributed scheduling, learning-based adaptations, and extensions to other AoI metrics.
%In this paper, we considered a VANET network, in which an RSU is responsible for scheduling the vehicles for broadcasting their updates as well as the updates of other vehicles. The graph of the network changes across time frames and the RSU, at the beginning of each frame, exploits the topology of the network graph as well as the AoI of vehicles to choose the sender vehicles as well as the target information they have to broadcast. We proposed the A-CB policy that indexes each target-sender pair by its achievable age-gain, i.e., the reduction in SAoI. We also proposed two scheduling algorithms called local TA-CB and global TA-CB that, in addition to the age-gains, exploit the number of second-hop neigbors and the betweenness centrality to weigh the links. Finally, we evaluated the proposed algorithms and showed 

%\bibliographystyle{./IEEEtran}
%\bibliography{./IEEEabrv,./IEEEexample}
%***************************************************************************************************
%********************* BIBLIOGARPHY *********************************************************
%***************************************************************************************************  
\setstretch{1}
\nocite{*}
\bibliographystyle{IEEEtran}
\bibliography{myref}

\end{document}